\documentclass[english,12pt,sort&compress]{iopart}
\usepackage[T1]{fontenc}
\usepackage[latin1]{inputenc}
\usepackage{float}
\usepackage{iopams}
\usepackage{setstack}
\usepackage{graphicx}
\usepackage{amssymb}
\usepackage[numbers]{natbib}
\usepackage{verbatim}
\usepackage{enumerate}

\makeatletter



\usepackage{babel}
\makeatother
\begin{document}

\title[]{Ensemble inequivalence : Landau theory and the ABC model}

\author{O Cohen and D Mukamel}

\address{Department of Physics of Complex Systems, Weizmann Institute of Science, Rehovot 76100, Israel}

\eads{\mailto{or.cohen@weizmann.ac.il}, \mailto{david.mukamel@weizmann.ac.il}}

\begin{abstract}
It is well known that systems with long-range interactions may exhibit different phase diagrams when studied within two different ensembles.
In many of the previously studied examples of ensemble inequivalence,
the phase diagrams differ only when the transition in one of the ensembles is first order.
By contrast, in a recent study of a generalized ABC model, the
 canonical and grand-canonical ensembles of the model were shown to differ even when they both exhibit a continuous transition.
Here we show that the order of the transition where ensemble inequivalence may occur is related to the symmetry properties of the order parameter associated with the transition.
 This is done by analyzing the Landau expansion of a generic model with long-range interactions.
 The conclusions drawn from the generic analysis are demonstrated for the ABC model by explicit calculation of its Landau expansion.
\end{abstract}


\noindent{\it Keywords\/}: Classical phase transitions (Theory), Phase diagrams (Theory), Driven diffusive systems (Theory)

\section{Introduction}

Much work has been devoted in recent years to the understanding of the properties of systems with long-range interactions.
These are systems where the two-body potential decays at large distance, $R$, as $1/R^{d-\sigma}$,
with $0\leq\sigma\leq d$ in $d$ dimensions. Interactions of this type are found
in various systems such as self-gravitating models \cite{Padmanabhan1990,Chavanis2002},
non-neutral plasmas \cite{Nickolson1992}, dipolar ferroelectrics \cite{LandauLifshitz1960} and
others.

The algebraic decay of the two-body potential results in an ill-defined thermodynamic limit since
the energy is super-extensive ($E \sim V^{1+\sigma/d}$, where $V$ is the volume of the system). A thermodynamic limit may be restored by
applying the Kac prescription \cite{Kac1963}, by which the potential strength is rescaled by
an appropriate volume-dependent factor. However, even within this scheme the energy remains nonadditive,
in a sense that the energy of a system composed of two subsystems is not necessarily equal to the sum of the energies of the two.
Nonadditivity results in some distinct properties that are not found
in the more commonly studied short-range interacting systems.
These properties include, for example, negative specific heat in micro-canonical ensembles (and similarly negative compressibility in
canonical ensembles) and inequivalence of statistical ensembles, which is the focus of this paper
\cite{Antonov1962,LyndenBell1968,Thirring1970,Hertel1971,LyndenBell1999,
Thirring2003,Barre2001,Barre2002,Mukamel2005, Misawa2006,Grosskinsky2008,Lederhendler2010a,Lederhendler2010b,Barton2010}.
Long-range interactions may also result in dynamical effects such as the existence of quasi-stationary states,
which are long-lived states different from the usual Gibbs-Boltzmann one \cite{Yamaguchi2003,Yamaguchi2004},
and anomalous relaxation time to equilibrium.
 Recent reviews of this topic are found in \cite{Dauxois2002,Campa2007,Mukamel2008,Campa2009,Dauxois2010}.

In many studies of ensemble inequivalence in models with long-range interactions, it was found that
whenever the two ensembles (either micro-canonical/canonical or canonical/grand-canonical) yield a continuous transition, the transition occurs at the same critical point.
 On the other hand, the phase diagrams of the two ensembles differ from each other when the phase transition becomes first order in one of the ensembles.
 Such behaviour has been observed in studies of numerous systems such as the spin-1 Blume-Emery-Griffiths (BEG) model \cite{Barre2002,Ellis2004,Touchette2004,Mukamel2005},
the ABC model \cite{Lederhendler2010a,Lederhendler2010b}, the
zero range process \cite{Grosskinsky2008}, two-dimensional vortices \cite{Venaille2007}, random field and spin glass systems \cite{Takahashi2011,Bertalan2011,Murata2012} and others.
However, general considerations have shown that different ensembles may, in principle, exhibit different critical points \cite{Bouchet2005}.
In fact, a recent study of a generalized ABC model revealed a case where
 the canonical and grand-canonical ensembles exhibit a second order phase transition at different critical points \cite{Barton2010}.

In this paper we consider a general framework within which one can relate the order of the phase transition where
ensemble inequivalence may be observed and
the symmetry properties of the underlying order parameter of the transition.
To this end, we consider a generic model which undergoes a phase transition governed by some order parameter, $m$. This parameter
could, for instance, be the average magnetization in the case
of a magnetic transition such as in the Ising model. We consider the model
 within a `higher' ensemble, where a certain thermodynamic variable, denoted by $q$, is allowed to fluctuate, and within a `lower' ensemble,
 where $q$ is kept at a fixed value.
 For instance, in the case where $q$ is the energy the two ensembles would correspond to the canonical and micro-canonical ensembles,
 whereas in the case where $q$ is the particle density they would correspond to the grand-canonical and canonical ensembles.

 A natural framework within which phase diagrams of systems with long-range interactions can be analyzed is the Landau theory.
 In this approach one considers the partition sum over the microscopic configurations of a system, $\mathcal{C}$,
 that correspond to some value of the thermodynamic variables, taken here to be $m$ and $q$, given by
\begin{equation}
\label{eq:landau_exp}
\exp( - \beta V f(m,q)) \equiv \sum_{\mathcal{C}\,:\,m(\mathcal{C})=m, q(\mathcal{C})=q} \exp (-\beta E(\mathcal{C})) .
\end{equation}
Here $E$ is the energy of a configuration, $V$ is the volume of the system and $f(m,q)$ is the large deviation function.
Landau theory is a phenomenological approach whereby $f(m,q)$ is assumed to be an analytic function even at criticality. This allows one to study
the phase diagram of the model by expanding $f(m,q)$ in powers of $m$ and $q$ around its minima.
In systems with long-range interactions, the Landau theory is expected to be exact to leading order in $1/V$.
This conjecture, whereby $f(m,q)$ is analytic,
is of course true for models with infinite-range mean-field type interactions, and has been
proven rigourously also for spin models with  $1/R^{d-\sigma}$ interactions for $0\leq\sigma\leq d$ under certain conditions \cite{Mori2010,Mori2011}.
We can therefore study the behaviour of long-range interacting models in a rather general context by
considering a generic Landau expansion of $f(m,q)$.
 Since the symmetry properties of $m$ and $q$ determine which
terms may appear in the Landau expansion, this approach allows us to study the connection between the symmetry of the model and the nature of its phase transitions.

In the present paper we show how the symmetry properties of $m$ and $q$ determine whether or not the two ensembles exhibit the same critical points.
 For simplicity we study here the case where $m$ and $q$ are scalars and where $f(m,q)$ is invariant under a simple transformation of $m$, of the form $m\to -m$.
Our results can be straightforwardly generalized to more elaborate cases, where for instance $m$ is a vector or a complex number, as demonstrated below in our study of the ABC model.
In the `lower' ensemble,
where $q$ is fixed, the Landau expansion of $f(m,q)$ around $m=0$ is given by
\begin{equation}
\label{eq:Fm0a}
f(m,q)=a(q)+c(q)m^{2}+e(q)m^{4} + O(m^6).
\end{equation}
The critical point of the `lower' ensemble is found where the coefficient of $m^2$ changes sign, and it is thus given by the equation $c(q)=0$.
In the `higher' ensemble, fluctuations in $q$ around its average value $q^\star$, denoted by $\delta q=q-q^\star$, may or may not affect the coefficient of $m^2$
depending on the symmetry of the model. In order to study this effect we expand the coefficients in \eref{eq:Fm0a} in powers of $\delta q$.
Two types of symmetries are considered in this expansion:
\begin{enumerate}
\item $f(m,q)=f(-m,q)$ : In this case the dominant coupling term between $m$ and $\delta q$ in the Landau expansion is of the form $m^2\delta q$.
As a result, near the critical point of the `higher' ensemble $\delta q$ scales as $\delta q \sim m^2$, implying that $\delta q$ affects the coefficient of the $m^4$ term in the Landau expansion and higher order terms.
Since the critical line is determined by the coefficient of $m^2$, the `lower' and the `higher' ensembles yield the same critical line.
\item $f(m,q)=f(-m,-q)$ : In this case we study the stability of phase where $q^\star =0$ for which this symmetry is relevant.
Here $f(m,q)$ may involve a bilinear term of the form $m \delta q$, which was not considered in \eref{eq:Fm0a}.
As a result, near the critical point of the `higher' ensemble $\delta q$ scales as $\delta q \sim m$. This scaling implies that $\delta q$ affects the coefficient of the $m^2$ term in the Landau expansion, causing the `higher' ensemble to exhibit a critical line which is different than that of the `lower' ensemble.
\end{enumerate}
The first symmetry, $f(m,q)=f(-m,q)$, corresponds to the more commonly studied case, where for instance $m$ is the average magnetization and $q$ is either the energy of the overall particle density.
 This is because both the energy and the particle density are invariant under any symmetry transformation of the model.

The conclusions drawn from the above generic analysis are demonstrated explicitly for the ABC model \cite{Evans1998,Evans1998b,Ayyer2009}.
This is a three-species exclusion model defined on a discrete one-dimensional lattice. The model exhibits a phase transition between a homogenous state and a phase-separated state, which is composed of three coexisting regions each predominantly occupied by one of the species.
Here we study the ABC model on an {\it interval}, with closed boundary conditions.
In this case, although the model evolves by local dynamics, its steady state is given
by the Gibbs-Boltzmann distribution of a system with long-range interactions \cite{Ayyer2009}.

The original ABC model evolves under particle-conserving dynamics where the overall density of each species, denoted by $r_A,r_B$ and $r_C$, is fixed.
Previous studies have introduced two kinds of particle-nonconserving generalizations of the ABC dynamics, corresponding to two distinct grand-canonical ensembles.
In the first generalization, referred to as {\it Model 1}, the overall density of the three species,
$r=r_A+r_B+r_C$, is allowed to fluctuate while the differences between their densities, $r_A-r_B$ and $r_B-r_C$, are fixed.
In the second generalization, referred to as {\it Model 2}, the differences, $r_A-r_B$ and $r_B-r_C$, are allowed to fluctuate while maintaining the overall particle density, $r$, constant.
Here we show that the two grand-canonical ABC models obey different symmetries, corresponding to the two different generic models mentioned above.
  In the ABC model the symmetry transformation corresponding to $m\to -m$ is a cyclic exchange
 of the labels of the three species, $A\to B$, $B\to C$ and $C\to A$. Since the fluctuating field of {\it Model 1}, $r=r_A+r_B+r_C$, is invariant under this transformation,
  {\it Model 1} and the canonical ABC model exhibit the same critical line, as shown previously in  \cite{Lederhendler2010a,Lederhendler2010b}.
 On the other hand, since the fluctuating fields of {\it Model 2}, $r_A-r_B$ and $r_B-r_C$, are not invariant under the label-exchange, {\it Model 2} and the canonical ABC model
 exhibit different critical lines, as shown previously in \cite{Barton2010}.
 By carrying out the Landau expansion of the two grand-canonical ABC models explicitly, we derive their critical points and demonstrate the two models are related
 to the generic models mentioned above. In addition, the Landau expansion of {\it Model 2} provides an alternative derivation of its critical point,
  computed previously in \cite{Barton2010} by deriving the exact expression of the steady-state density profile of {\it Model 2}.

The paper is organized as follows: In \sref{sec:landau} we study ensemble inequivalence in the context of two generic models with long-range interactions,
each exhibiting a different symmetry, as discussed above.
 The results of \sref{sec:landau} are demonstrated on the ABC model, which is presented in \sref{sec:can_ABC}. We study the two grand-canonical generalizations of the ABC model, {\it Model 1} and {\it Model 2}, by explicitly computing their Landau expansion in sections \ref{sec:ABC1} and \ref{sec:ABC2}, respectively. Concluding remarks are presented in \sref{sec:conc}

\section{Ensemble inequivalence in a generic Landau model}
\label{sec:landau}

\subsection{Symmetry of the type $f(m,q)=f(-m,q)$}
\label{sec:landau_scalar}
In this section we consider a generic model whose Landau free energy is invariant under the
transformation $(m,q)\to(-m,q)$, such as in the case where $m$ and $q$ correspond to the overall magnetization and
the energy of the system, respectively. We study the phase diagram of the generic model within a lower ensemble, where $q$ is fixed, and within a higher
ensemble where $q$ is allowed to fluctuate, and show that the two ensembles exhibit the same critical line.

\paragraph{Lower ensemble: }
In the lower ensemble, where $q$ is fixed, and close to the transition temperature where $m$ is small,
the Landau free energy density can be expanded in powers of $m$ as
\begin{equation}
\label{eq:Fm0}
f(m,q)=a_{0}(q)+c_{0}(q)m^{2}+e_{0}(q)m^{4} + O(m^6).
\end{equation}
The above coefficients are labeled with $0$ subscript in order to be consistent with their expansion in
small fluctuations in $q$, carried out below in the study of the higher ensemble.
The free energy of the lower ensemble is obtained by minimizing the Landau free energy with respect to $m$,
\begin{equation}
F(T,q,\dots) = \min_m f(m,q) = f(m^\star,q),
\end{equation}
where $m^\star$ is the average magnetization for a given value of $q$. Here $T$ is the temperature and the dots denote other
possible parameters of the system. In order simplify the notation we omit here and below the dependence of the
coefficients of the Landau expansion on $T$ and those additional parameters.

\begin{figure}[t]
\noindent
\begin{centering}\includegraphics[scale=1.0]{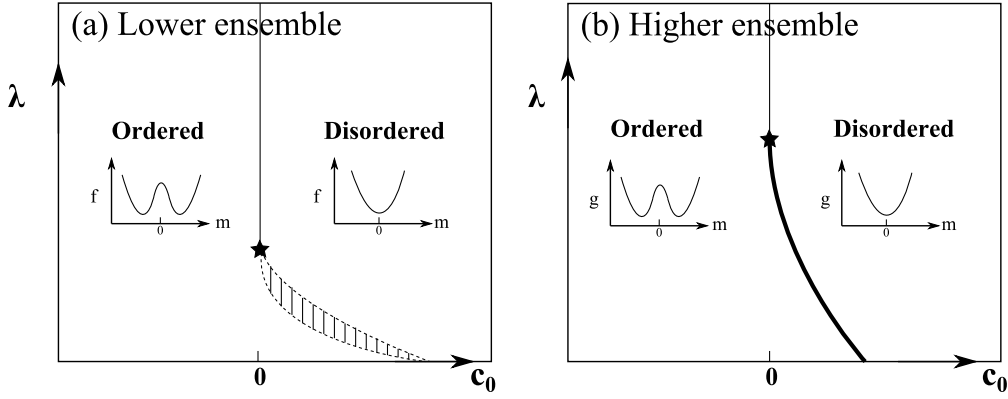}\par\end{centering}
\caption{ Sketch of the phase diagram of the lower (a) and higher (b) ensemble of the generic model where $f(m,q)=f(-m,q)$.
Both diagrams show the same second order transition line (thin solid line) which terminates at a different tricritical point ($\star$) in (a) and in (b), where
the transition becomes first order. The first order transition is denoted by a thick solid line in (b) and by a discontinuity in $\lambda$ in (a) (vertical lines).
\label{fig:generic_diagram1}}
\end{figure}

The phase diagram of the lower ensemble can be derived from \eref{eq:Fm0} as follows:
At $c_0=0$ and for $e_0>0$, the model undergoes a second order transition between a disordered phase, for which $c_0>0$ and $m^\star=0$, and an ordered phase, where $c_0<0$ and $m^\star\neq0$.
This transition line is denoted by the thin solid line in \fref{fig:generic_diagram1}a.
The figure shows a sketch of the phase diagram of the model in the $(c_0,\lambda)$ plane, where $\lambda$ is the conjugate field of $q$, given by
 \begin{equation}
 \lambda(q)=df(m^\star,q)/dq.
 \end{equation}
 On the second order transition line $\lambda(q)$ changes smoothly. This transition line terminates at tricritical point ($\star$), where $c_0=e_0=0$.
For $e_0<0$, the critical line is preempted by a first order phase transition, taking place at $c_0>0$, whose value is determined by the higher order terms in the Landau expansion.
The first order phase transition is denoted by the vertical lines in \fref{fig:generic_diagram1}a, which mark the discontinuity in $\lambda(q)$, observed as the system switches between a phase where $m^\star=0$, and a phase with a nonvanishing value of $m^\star$.

\paragraph{Higher ensemble: }
In the higher ensemble,  where $q$ is allowed to fluctuate,
the Landau free energy density is given by the Legendre transform of $f(m,q)$, defined as
\begin{equation}
\label{eq:Gm1}
 g(m,\lambda) = \min_q \left[ f(m,q)-\lambda q\right].
 \end{equation}
The free energy of the higher ensemble is obtained by minimizing $f(m,q)-\lambda q$ with respect
to $q$ and $m$ while keeping $\lambda$ fixed,
\begin{equation}
G(T,\lambda,\dots) = \min_{m,q} \left( f(m,q)-\lambda q \right) = f(m^\star,q^\star)-\lambda q^\star.
\end{equation}
Close to critical line we consider small fluctuations around $q^\star$, denoted by $q=q^\star+\delta q$, which yields the following expansion,
\begin{eqnarray}
\label{eq:Fdq}
f(m,q) -\lambda q & =& \left[ a_0(q^\star)-\lambda q^\star \right]  + a_2(q^\star)(\delta q)^2 \\ &+ &  \left[ c_0(q^\star) + c_1(q^\star) \delta q +c_2(q^\star)(\delta q)^2 \right]m^2+e_{0}(q^\star)m^{4}+ O(m^6) \nonumber ,
\end{eqnarray}
where $\alpha_n (q) = \frac{1}{n!}\frac{\partial^n\alpha_0}{\partial q^n} $ for $\alpha=a,c,e$.
Here, we omitted the linear term in $\delta q$ since \eref{eq:Fdq} is minimal at $q=q^\star$ and $m=0$, which implies that
\begin{equation}
 \label{eq:q_of_lambda}
a_1(q^\star) =\lambda q^\star.
\end{equation}
In order to obtain the Landau expansion of the higher ensemble we minimize \eref{eq:Fdq} with respect to $\delta q$, yielding
\begin{equation}
\label{eq:dqm2}
\delta q = - \frac{c_1(q^\star)}{2a_2(q^\star)} m^2 + O(m^4).
\end{equation}
The scaling of the fluctuations in $q$, $\delta q \sim m^2$, is a direct consequence of the
symmetry of $f(m,q)$, by which the lowest order coupling term between $m$ and $\delta q$ in $f(m,q)$ is of the form  $m^2 \delta q$.
This scaling implies that the fluctuations in $q$ affect only the coefficient of $m^4$ and
 higher order terms in $g(m,\lambda)$.
The Landau expansion of the higher ensemble is obtained by inserting \eref{eq:dqm2} into \eref{eq:Fdq} and the result into \eref{eq:Gm1}, yielding
\begin{equation}
\label{eq:gmlambda1}
g(m,\lambda) =  c_0(q^\star) m^2 + \left[ e_{0}(q^\star) -
 \frac{c_1^2(q^\star)}{4a_2(q^\star)}\right] m^4 + O(m^6).
 \end{equation}
Here the dependence on $\lambda$ is expressed through $q^\star(\lambda)$, given by \eref{eq:q_of_lambda}.
The higher ensemble therefore exhibits the {\it same} critical line as the lower
ensemble, at $c_0=0$. Here, however, the second order transition line terminates at a {\it different} tricritical point, given by $e_0=c_1^2/4a_2$. A sketch of the resulting phase diagram is shown in \fref{fig:generic_diagram1}b.
The plot shows a second order transition line (thin solid line) which turns into a first order (thick solid line) at a tricritical point ($\star$), which is different than the tricritical point of the lower ensemble.

\Fref{fig:generic_diagram1} corresponds to the case where the coefficient of the $m^6$ term in $f(m,q)$ and $g(m,\lambda)$ is positive,
and thus the second and the first order transition lines join at a tricritical point in both ensembles.
When the coefficient of $m^6$ is negative the two lines join at a critical-end-point, yielding
a different phase diagram which will not be discussed here. An example of a critical-end-point was found in a variant of {\it Model 1} in \cite{Lederhendler2010b}.

\subsection{Symmetry of the type $f(m,q)=f(-m,-q)$}
\label{sec:landau_tensor}
In this section we consider a case where the generic Landau free energy has a symmetry of the form $f(m,q)=f(-m,-q)$.
A simple example of this symmetry is provided by the anisotropic XY model with infinite-range interactions, as discussed at the end of this section.
This type of symmetry is also relevant for {\it Model 2} studied in \sref{sec:ABC2}.
As in the previous section we study the generic model within a lower and a higher ensemble, which in this case yield different critical points.

\begin{figure}[t]
\noindent
\begin{centering}\includegraphics[scale=1.0]{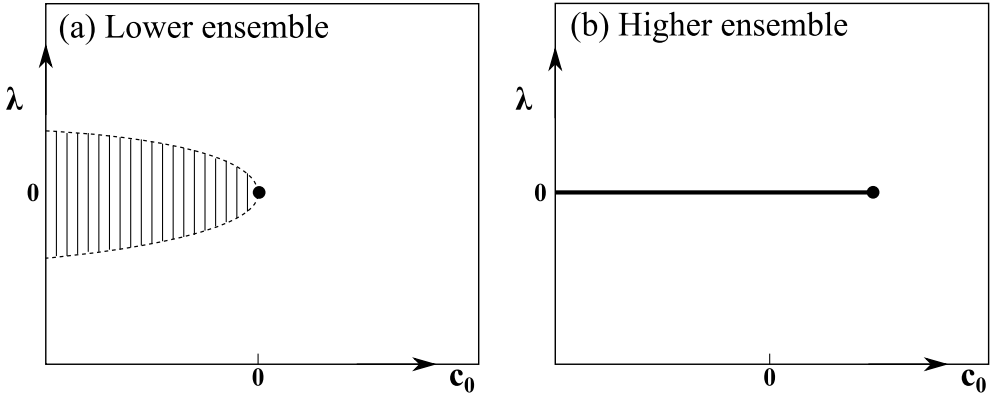}\par\end{centering}
\caption{ Sketch of the phase diagram of the lower (a) and higher (b) ensemble of the generic model where
$f(m,q)=f(-m,-q)$. Here the two ensembles display different critical points ($\bullet$) at $c_0=0$ in (a)
and at $c_0=b_1^2/4a_2$ in (b). For lower values of $c_0$ we find a first order transition,
denoted in (b) by a solid line and in (a) by a discontinuity in
$\lambda$ (vertical lines).
\label{fig:generic_diagram2}}
\end{figure}

\paragraph{Lower ensemble: }
In the lower ensemble, where $q$ is fixed, and close to the transition temperature, we consider a general Landau expansion of the form
\begin{equation}
\label{eq:Fm1}
f(m,q)=a_{0}(q)+b_{0}(q)m+c_{0}(q)m^{2} +d_{0}(q)m^{3}+ e_{0}(q)m^{4}+ O(m^5).
\end{equation}
In order to satisfy the symmetry, $f(m,q)=f(-m,-q)$, the coefficients of the even powers of $m$ must obey
\begin{equation}
\label{eq:Fm3}
a_0(q)=a_0(-q) \quad c_0(q)=c_0(-q) \quad  e_0(q)=e_0(-q),
\end{equation}
while coefficients of the odd powers of $m$ obey
\begin{equation}
\label{eq:Fm3a}
b_0(q)=-b_0(-q), \quad d_0(q)= - d_0(-q).
\end{equation}
We study the model close to the $q=0$ line, where \eref{eq:Fm3} and \eref{eq:Fm3a} play a role in determining which terms appear in the Landau expansion of the higher ensemble, as discussed below.
For $q=0$ all the coefficients of the odd powers of $m$ in \eref{eq:Fm1} vanish, and the model exhibit a critical point at $c_0(0)=0$ (note that $c_0$ depends on $T$ and possibly other parameters that
are not mentioned here explicitly).
  The order of the transition at this point depends on the sign of $e_0(0)$. The phase diagram of the lower ensemble
  for the case of $e_0(0)>0$, where the transition
is of second order,  is sketched in \fref{fig:generic_diagram2}a.
It is easy to see from
 \eref{eq:Fm1}-\eref{eq:Fm3a} that above the transition, for $c_0>0$, the $q=0$ line
corresponds to $\lambda(q)=df(m^\star,q)/dq=0$. Below the critical point ($\bullet$), namely for $c_0<0$, the lower ensemble undergoes a first order transition at $q=0$, in which
the conjugate field of $q$ exhibits a discontinuity as the system switches between a phase with a positive nonvanishing $m^\star$ and a phase  with a negative nonvanishing $m^\star$ (vertical lines).

\paragraph{Higher ensemble: }
In the higher ensemble, where $q$ is allowed to fluctuate, we follow the same lines of derivation presented in \sref{sec:landau_scalar}.
We consider the expansion of  $f(m,q)-\lambda q$ in powers of $m$ and $\delta q$ around $q^\star=0$, which is given by
\begin{eqnarray}
\label{eq:Fdq2}
f(m,\delta q) = a_0(0) + a_2(0)(\delta q)^2  +   b_1(0) m \delta q  + c_0(0)m^2+ O(m^4).
\end{eqnarray}
 By minimizing $f$ with respect to $\delta q$ we obtain to lowest order
\begin{equation}
\label{eq:Fdq2b}
\delta q = - \frac{b_1(0)}{2 a_2(0)} m+O(m^2).
\end{equation}
This linear scaling, $\delta q \sim m$, is a direct consequence of the
symmetry of $f(m,q)$, which allows for a bilinear coupling term of the form $m \delta q$.
As a result of this scaling, the fluctuations in $q$ affect the coefficient of $m^2$ in $g(m,\lambda)$.
Inserting \eref{eq:Fdq2b} into \eref{eq:Fdq2} and the result into \eref{eq:Gm1} yields
\begin{equation}
\label{eq:Gm2}
 g(m,\lambda) = a_0(0) +  \left[ c_{0}(0) -
 \frac{b_1^2(0)}{4a_2(0)}\right] m^2 + O(m^4).
 \end{equation}
We therefore conclude that the higher ensemble has a critical point at $c_{0}=b_1^2/4a_2$, which is different
from the one found in the lower ensemble, given by $c_0=0$. The order of the transition at the critical
point depends on the higher order terms in \eref{eq:Gm2}, which are not considered here explicitly.
The resulting phase diagram of the higher ensemble is sketched in \fref{fig:generic_diagram2}b for the
case where the coefficient of $m^4$ in \eref{eq:Gm2} is positive at the critical point. The figure
shows the critical point ($\bullet$), below which the higher ensemble undergoes a first order phase
transition at $\lambda=0$. Here of course, since $\lambda$ is the control parameter, it does not exhibit a discontinuity.

A simple example for a model that obeys the symmetry discussed above is the anisotropic XY model with
infinite-range interactions. The model is defined on a one-dimensional lattice of $N$ sites. The equilibrium state of the model is a Gibbs-Boltzmann distribution corresponding to the following Hamiltonian,
\begin{equation}
H =- \frac{J_{xx}}{2N} \left(\sum_{i=1}^N S_{i,x} \right)^2 -\frac{J_{yy}}{2N} \left(\sum_{i=1}^N S_{i,y} \right)^2 - \frac{J_{xy}}{2N} \left(\sum_{i=1}^N S_{i,x}\sum_{j=1}^N S_{i,y} \right).
\end{equation}
Here $S_{i,x},S_{i,y}=\pm 1$ are the $x$ and $y$ components of the spin at site $i$, respectively, and $J_{xx} \geq J_{yy}>0$.
By defining the average magnetization in each direction as $m_\alpha=N^{-1}\sum S_{\alpha,i}$ for $\alpha=x,y$
we obtain a Landau free energy density of the form
\begin{eqnarray}
\label{eq:XYf}
f(m_x,m_y)&= & -\frac{1}{2} \left( J_{xx} m_x^2+ J_{xy} m_x m_y + J_{yy} m_y^2\right) \\
      &+& T  \sum_{\alpha=x,y} \left[ \frac{1+m_\alpha}{2}\ln(\frac{1+m_\alpha}{2})+\frac{1-m_\alpha}{2}\ln(\frac{1-m_\alpha}{2})\right],\nonumber
\end{eqnarray}
where we took $k_B=1$ in order to simplify the notations.

We study the model within a lower ensemble,  where $m_y$ is fixed, and within a higher ensemble where $m_y$ is allowed to fluctuate.
From the symmetry of \eref{eq:XYf}, whereby $f(m_x,m_y)=f(-m_x,-m_y)$, we expect to find a phase diagram similar to \fref{fig:generic_diagram2},
with $m_x$ and $m_y$ playing the role of $m$ and $q$, respectively. 
In the lower ensemble we follow the lines of derivation presented for the generic model and obtain a critical point at $m_y=0$ and
\begin{equation}
\label{eq:Tc_heis_low}
 T_c^{LE}= J_{xx}.
\end{equation}
In the higher ensemble, the Landau free energy is given by
\begin{equation}
g(m_x,H_y)=\min_{m_y}\left[f(m_x,m_y)-H_y m_y\right],
\end{equation}
where $H_y$ is the magnetic field in the $y$ direction, corresponding to field $\lambda$ of the generic model.
By analyzing the Landau expansion of the model around $m_x=m_y=0$ and $H_y=0$,
 we find that a specific linear combination of $m_x$ and $m_y$ becomes unstable at
\begin{equation}
\label{eq:Tc_heis_high}
 T_c^{HE}= \frac{1}{2}\left( J_{xx}+J_{yy}+\sqrt{(J_{xx}-J_{yy})^2 +  J_{xy}^2}\right).
\end{equation}
The resulting phase diagram is similar to the one sketched in \fref{fig:generic_diagram2}. As expected, due to the symmetry of the model, whereby $f(m_x,m_y)=f(-m_x,-m_y)$,
the lower and the higher ensembles exhibit different second order transition points given by \eref{eq:Tc_heis_low} and \eref{eq:Tc_heis_high}, respectively.

\begin{figure}[t]
\noindent
\begin{centering}\includegraphics[scale=0.7]{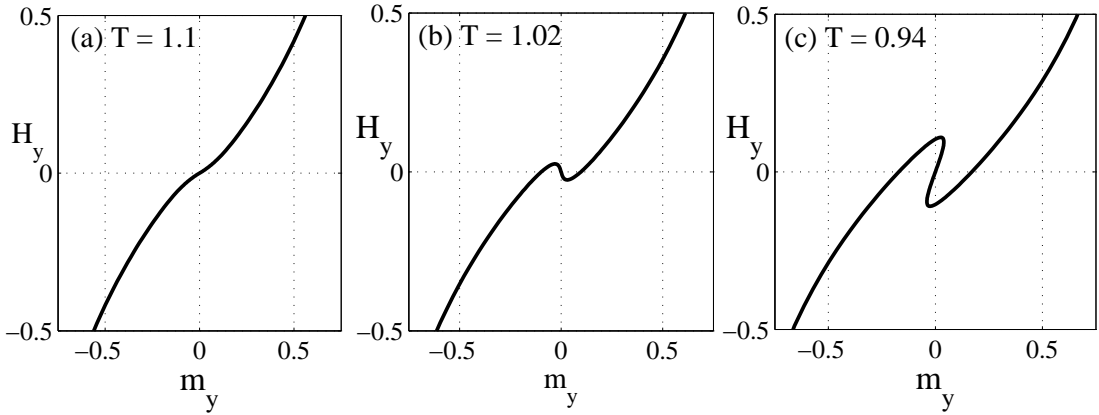}\par\end{centering}
\caption{
The magnetic field in the $y$ direction in the XY model, $H_y=df(m^\star_x,m_y)/d m_y$, as a function $m_y$ for $J_{xx}=1,J_{yy}=0.5$ and $J_{xy}=0.1$. For this choice of parameters $T^{LE}_c=1$ and $T^{HE}_c\approx 1.06$.
In (a), where $T>T^{HE}_c$, the one-to-one correspondence between $H_y$
and $m_y$  implies that the ensembles are equivalent. In (b), where $T^{LE}_c<T<T^{HE}_c$, since $H_y$ becomes nonmonotonous only the higher ensemble undergoes a first order phase transition at $H_y=0$, where it exhibits a discontinuity in $m_y$. For $T<T^{LE}_c$, shown in (c), the lower ensemble also undergoes a first order phase transition at $m_y=0$, where it exhibits a discontinuity in $H_y$.

\label{fig:my_lambda}}
\end{figure}

Below the critical point, the XY model undergoes a first order phase transition. The nature of this transition can be understood by
plotting the magnetic field in the $y$ direction in the lower ensemble, given by $H_y=df(m^\star_x,m_y)/d m_y$, where ${m}_x^\star$ the solution of
 $df(m_x,m_y)/dm_x=0$.
In \fref{fig:my_lambda}a, where $T>T^{HE}_c$, we find a one-to-one correspondence between $H_y$
and $m_y$ which implies that the two ensembles are equivalent.
 For lower temperatures, in \fref{fig:my_lambda}b where $T^{LE}_c<T<T^{HE}_c$, since $H_y$ becomes nonmonotonous the higher ensemble undergoes a first order phase transition at $H_y=0$, signified by a discontinuity in $m_y^\star(H_y)$. On the other hand, the lower ensemble does not undergo any phase transition in \fref{fig:my_lambda}b.
  For $T<T^{LE}_c$, shown in \fref{fig:my_lambda}c, the lower ensemble also exhibits a first order phase transition as we vary $m_y$ through $m_y=0$, yielding a discontinuity in $H_y$.

\section{ABC model on an interval}
\label{sec:ABC_model}
The general discussion presented in the previous section may be applied to the ABC model.
 This is a three-species exclusion model studied here on an {\it interval}, where its dynamics obeys detailed balance with respect to a Hamiltonian with long-range interactions.
In \sref{sec:can_ABC} we review the phase diagram of the canonical ABC model, where the overall densities of each of the three species are conserved.
We then study two distinct grand-canonical generalizations of
the model in sections \ref{sec:ABC1} and \ref{sec:ABC2}, referred to as {\it Model 1} and {\it Model 2} , respectively, where different conservation rules are relaxed. By explicitly calculating their
 Landau free energy, {\it Model 1} and {\it Model 2} are shown to correspond to the generic models presented in sections \ref{sec:landau_scalar} and \ref{sec:landau_tensor}, respectively.

\subsection{ABC model on an interval: canonical ensemble}
\label{sec:can_ABC}
The ABC model is a one-dimensional driven exclusion model which
has been studied extensively in recent years, mainly as an example of a model which, when studied
on a ring geometry, exhibits long-range order and a phase-separation transition out of equilibrium \cite{Evans1998,Evans1998b,Clincy2003}.
Here we consider the model on an {\it interval}, where its dynamics satisfies detailed balance \cite{Ayyer2009}.
Each of the $L$ lattice sites can be occupied by one particle (either $A$,$B$ or $C$) or remain empty.
The microstate is thus denoted by ${\boldsymbol \zeta }=\left\{{ \zeta}_i\right\}_{i=1}^{L}$, where $i$ is the index of the site and $\zeta_i=A,B,C$ or $0$.
 The numbers of $A$, $B$ and $C$ particles are denoted by $N_{A},\, N_{B}$ and $N_{C}$, respectively, with $N_{A}+N_{B}+N_{C}=N \leq L$.
 We also denote the average density of each species by $r_\alpha=N_\alpha/L$ for $\alpha=A,B,C$ and the overall particle density by $r=r_A+r_B+r_C\leq 1$.
The model evolves by random sequential updates whereby particles on neighbouring sites are exchanged with the following rates,
\begin{equation}
\label{eq:abc_rules}
AB\overset{q}{\underset{1}{\rightleftarrows}}BA,\qquad BC\overset{q}{\underset{1}{\rightleftarrows}}CB,\qquad CA\overset{q}{\underset{1}{\rightleftarrows}}AC.
\end{equation}
 The vacancies are inert and exchange symmetrically with the other particles with the following rates,
\begin{equation}
\label{eq:vacancyexchange}
A0\overset{1}{\underset{1}{\rightleftarrows}}0A,
\qquad B0\overset{1}{\underset{1}{\rightleftarrows}}0B,
\qquad C0\overset{1}{\underset{1}{\rightleftarrows}}0C.
\end{equation}
For $q=1$ the model relaxes to a homogeneous steady state where the particles are randomly
distributed on the lattice, whereas for $q\neq 1$ the particles phase separate
into three domains in the limit $L\to\infty$. The vacancies remain homogenously distributed in the
system for any value of $q$.
Without loss of generality we consider here $q<1$ where the domains are arranged clockwise as $AAA \dots BBB \dots CCC$, with vacancies distributed homogenously on the lattice.
A unique feature of the ABC model on an interval is that although its dynamics is local, the model obeys detailed balance with respect to an effective Hamiltonian with long-range interactions, given by
\begin{equation}
\label{eq:abc_ham}
\mathcal{H}\left(\boldsymbol \zeta \right)=\sum_{i=1}^{L-1}\sum_{j=1}^{L-i}\left(A_{i}C_{i+j}+B_{i}A_{i+j}+C_{i}B_{i+j}\right)-\frac{1}{6}N^2.
\end{equation}
The projection operators in the Hamiltonian are defined as
\begin{eqnarray}
\alpha_{i}\left(\boldsymbol \zeta \right)= \left\{\begin{array}{ccc}
1 & \: & \mathrm{for} \,\, \zeta_{i}=\alpha \\
0 & \: & \mathrm{else} \end{array} \right. ,
\end{eqnarray}
for $\alpha=A,B$ and $C$.
The fact that a particle on site $i$ interacts equivalently with all the particles to its
right (on sites $i+j$ with $j=1,\dots, L-i$) implies that $\mathcal{H}\left(\boldsymbol \zeta \right)$ is a long-range Hamiltonian.
For convenience, we add a constant term to the Hamiltonian, $-\frac{1}{6}N^2$, which sets the average energy of the disordered phase to zero.
The fact that this terms scales super-linearly with the size of the system is another evidence for the long-range character of the ABC model.
On a ring geometry, not studied here, the model obeys detailed balance with respect to \eref{eq:abc_ham} only for $N_A=N_B=N_C$, whereas on an interval
this is true for arbitrary $N_A,N_B$ and $N_C$.

The ABC model is often studied in the limit of weak asymmetry, where $q$ approaches $1$ in the thermodynamic limit as $q=e^{-\beta/L}$. Here the parameter $\beta$ serves as the inverse temperature of the model.
For equal densities, $N_A=N_B=N_C$, the model undergoes a second order transition in the $L\to \infty$ limit between a disordered phase and an ordered phase, with
three macroscopic domains each predominantly occupied by one of the species. The transition point, given by
\begin{equation}
\label{eq:abc_bc}
\beta_{c}=2\pi\sqrt{3}/r,
\end{equation}
has been derived for $r\equiv N/L =1$ in \cite{Clincy2003, Ayyer2009} and for $r\leq1$ in \cite{Lederhendler2010a}. For nonequal densities, the model relaxes to an ordered steady-state profile for all values of $\beta$, and
hence no phase transition takes place \cite{Ayyer2009}.
The properties of the weakly-asymmetric ABC model close to its transition point has been the focus of a number of recent studies,
where the authors investigated its steady-state correlations \cite{Bodineau2008,Bodineau2011,Gerschenfeld2012} as well as its dynamical properties \cite{Bertini2011,Gerschenfeld2011}.

In the rest of this section we derive the critical line of the canonical ABC model following
the lines presented in \cite{Lederhendler2010b}. We use, however, a different notation
which is found to be convenient for analyzing the grand-canonical generalizations of the ABC model presented in the following sections.

The weakly-asymmetric ABC model can be studied in the hydrodynamic limit where the probability to observe a particle of type $\alpha$ at site $i$ is represented by a continuous density profile, $\rho_\alpha(x)$,
with $x=i/L\in [0,1)$ \cite{Evans2000,Fayolle2004,Ayyer2009}. The canonical large deviation function of the hydrodynamic profile is given by
\begin{eqnarray}
\label{eq:abc_energy}
\mathcal{F}\left[{\boldsymbol \rho}(x),r\right] &= &\int_{0}^{1}dx\sum_\alpha\rho_{\alpha}(x)\ln\rho_{\alpha}(x) + \left(1-r\right)\ln\left(1-r\right)\nonumber \\
&+&\beta\left(\int_{0}^{1}dx\int_{0}^{1-x}dz\sum_\alpha\rho_{\alpha}(x)\rho_{\alpha-1}(x+z)-\frac{1}{6}r^{2}\right),
\end{eqnarray}
were $\boldsymbol \rho \equiv (\rho_A, \rho_B,\rho_C)$ and the index $\alpha$ runs cyclically over $A,B$ and $C$. \Eref{eq:abc_energy} implies that the
probability to observe a density profile, ${\boldsymbol \rho}(x)$, is given by
$P[\boldsymbol {\rho}(x)]\sim e^{-L\mathcal{F}\left[{\boldsymbol \rho}(x),r\right]}$.
 The first integral in \eref{eq:abc_energy}
is the entropy of the density profile (the logarithm of the number of microstates corresponding to ${\boldsymbol \rho}$), and the second integral is the continuum limit of the Hamiltonian \eref{eq:abc_ham}.
In order to find the critical point of the model we expand the density profile
around the equal-densities homogeneous profile in terms of the following Fourier modes,
\begin{equation}
{\boldsymbol \rho}(x)=\frac{r}{3} \left(\begin{array}{c}
1\\
1\\
1\end{array}\right)+\sum_{n=-\infty}^{\infty}{\bf a}_{n}e^{2\pi inx},\label{eq:abc_fourier}
\end{equation}
where $a_0=0$ and ${\bf a}_{n}=(a_{n,A},a_{n,B},a_{n,C})$ is a complex vector which obeys  ${\bf a}_n={\bf a}_{-n}^\star$ in order to assure that ${\boldsymbol \rho}(x)$ is real.
Here and below the $\star$ superscript denotes the complex conjugate.
Since we study the model close to
the homogenous phase we assume $|a_{n,\alpha}|\ll1$. The lowest order terms in $a_{n,\alpha}$ of the Landau expansion of \eref{eq:abc_energy} are given by
\begin{eqnarray}
\label{eq:abc_energy1}
\mathcal{F}(\{{\bf a}\}) &= &\mathcal{F}_{h}(r) \\
&+& \sum_{n\neq 0 }
{\bf a}_{n}^T\left[ \frac{3}{r}I-\frac{i\beta}{2\pi n}\left(\begin{array}{ccc}
0 & 1 & -1\\
-1 & 0 & 1\\
1 & -1 & 0\end{array}\right)\right] {\bf a}_{-n} + O(a_{n,\alpha}^4)\nonumber
,
\end{eqnarray}
where the ${\bf a}_{n}^T$ denotes the transposed vector of ${\bf a}_{n}$, $I$ denotes the identity matrix,
and $\mathcal{F}_{h}(r)=r\ln(r/3)+(1-r)\ln(1-r)$ is the free energy density of the homogenous phase multiplied by $\beta$.

\begin{figure}[t]
\noindent
\begin{centering}\includegraphics[scale=1.2]{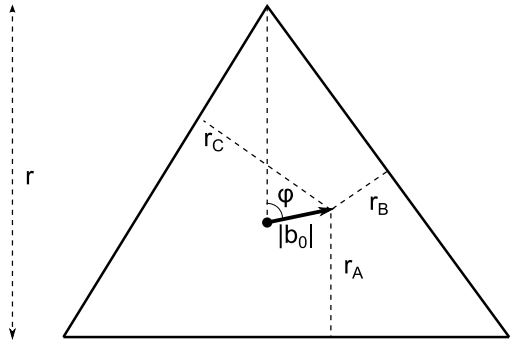}\par\end{centering}
\caption{Geometric representation of the amplitude $b_0$ which measures the
deviation from equal densities. The angle in the sketch is given by $\varphi= \mathrm{arg}(b_0)+\pi/6$. The lengths of the dashed
lines perpendicular to the edges denote the average densities of the three species, given by \eref{eq:b0_interp}.
\label{fig:b0}}
\end{figure}

The stability analysis of the Fourier modes can be carried out more conveniently by first diagonalizing the lowest order terms of $\mathcal{F}(\{{\bf a}\})$.
To this end we express the density profile in terms of the eigenvectors of the matrix which appears inside the square brackets in \eref{eq:abc_energy1} given by
\begin{eqnarray}
\label{eq:abc_vectors}
{\bf u}=\left(\begin{array}{c}
1\\
1\\
1\end{array}\right), \quad
{\bf v}=\frac{1}{\sqrt{3}}\left(\begin{array}{c}
1\\
e^{-2\pi i/3}\\
e^{2\pi i/3}\end{array}\right), \quad
{\bf v}^\star=\frac{1}{\sqrt{3}}\left(\begin{array}{c}
1\\
e^{2\pi i/3}\\
e^{-2\pi i/3}\end{array}\right).
\end{eqnarray}
The profile can thus be written as
\begin{eqnarray}
\label{eq:rho_expand}
{\boldsymbol \rho}(x)=\frac{r}{3}{\bf u}+\sum_{n=-\infty}^\infty\left[b_{n}{\bf v}+b_{-n}^\star{\bf v}^\star
\right]e^{2\pi inx},
\end{eqnarray}
where $b_n$ are the complex coefficients of the Fourier expansion.
Since the steady-state profile of the inert vacancies is flat, namely $\rho_{0}(x)=1-\rho_{A}(x)-\rho_{B}(x)-\rho_{C}(x)=1-r$ is independent of $x$,
we omitted from \eref{eq:rho_expand} the amplitudes of the vector ${\bf u} e^{2\pi i n x}$ for $n\neq 0$.
By integrating \eref{eq:rho_expand} over $x$, $b_0$ can be shown to set the deviation from equal-densities point through
\begin{equation}
\label{eq:b0_interp}
(r_A,r_B,r_C)=(\frac{r}{3},\frac{r}{3},\frac{r}{3})+|b_0|( \cos\varphi,\cos(\varphi+\frac{2\pi}{3}),\cos(\varphi-\frac{2\pi}{3})),
\end{equation}
where $\varphi=\mathrm{arg}(b_0)+\pi/6$, as depicted in \fref{fig:b0}. By summing the vector components in \eref{eq:b0_interp} we see that
$b_0$ controls the differences between the densities, $r_A-r_B$ and $r_B-r_C$, but not the overall particle density, which is given by $r=r_A+r_B+r_C$.
Since the canonical model undergoes a phase transition only for equal densities, we assume in this section
that $b_0=0$.

To lowest order in ${b_n}$ the canonical Landau expansion is given by
\begin{equation}
\label{eq:abc_energy2}
\mathcal{F}(\{{b}\})=\mathcal{F}_{h}(r)+\sum_{n\neq0}\left[(\frac{3}{r}-\frac{\sqrt{3}\beta}{2\pi n})|b_{n}|^{2}\right]+O(|b_{n}|^{4}).
\end{equation}
 At $\beta_c=2\pi\sqrt{3}/r$, the first mode, $b_{1}$, becomes unstable,
whereas the other modes are linearly stable.
Minimizing the free energy with respect to $b_n$ close to the transition point, $\beta-2\pi\sqrt{3}/r \ll 1$, yields to lowest order
\begin{equation}
\label{eq:abc_b_scaling}
b_{1+ 3l} \sim b_1^{1+3l} \qquad b_{1-3l} \sim (b_1^\star)^{|1-3l|},
\end{equation}
for $l=1,2,\dots$ This scaling is a result of the third order coupling terms in $\mathcal{F}$ of the form $b_{n_1}b_{n_2}b_{-n_1-n_2}$ and $b_{n_1}^\star b_{n_2}^\star b_{-n_1-n_2}^\star$. The other modes vanish,
\begin{equation}
\label{eq:abc_b_scaling1}
\qquad b_{\pm 3l} = 0 \qquad b_{-1\pm 3l}=0,
\end{equation}
since they are driven by $b_{-1}$ which is not excited.
 The ABC model thus has a complex primary order parameter, $b_{1}$, which drives some of the other Fourier modes.
The analysis of Landau expansion of the ABC model thus requires a strait-forward generalization of the generic canonical analysis presented in \sref{sec:landau_scalar}, where the primary order parameter, $m$, is real. By expressing the higher order Fourier modes in terms of $b_1$ and inserting the
result back into \eref{eq:abc_energy2} we obtain the Landau expansion of the model in terms $|b_{1}|$ alone, given by
\begin{equation}
\label{eq:can_abc_F}
\mathcal{F}(b_{1})=\mathcal{F}_{h}(r)+f_{2}\left(\beta,r\right)|b_{1}|^{2}+f_{4}\left(\beta,r\right)|b_{1}|^{4}+\mathcal{O}\left(|b_{1}|^{6}\right),
\end{equation}
where
\begin{equation}
f_{2}\left(\beta,r\right)=\frac{3}{r}-\frac{\sqrt{3}\beta}{2\pi},\qquad f_{4}\left(\beta,r\right)=\frac{9}{2r^{3}}\left(\frac{\sqrt{3}\beta r+6\pi}{\sqrt{3}\beta r+12\pi}\right)>0.
\end{equation}
The full derivation of $f_2$ and $f_4$ is found in \cite{Lederhendler2010b}.
We can now compare \eref{eq:can_abc_F} with the generic Landau expansion, found in \eref{eq:Fm0},
and derive the phase diagram of the canonical ABC model following the lines presented for the generic case in \sref{sec:landau_scalar}.
 Since $f_{4}>0$, we conclude that the canonical model exhibits only a second order transition line defined by the equation $f_2=0$, which yields
 \begin{equation}
\label{eq:bc1}
 \beta_{c}=2\pi\sqrt{3}/r.
\end{equation}

In the following two sections we study the phase diagrams of two different
grand-canonical generalizations of the ABC model. In \sref{sec:ABC1}, {\it Model 1} is shown to exhibit the same second order transition line as in \eref{eq:bc1}.
However, in contrast to the canonical ensemble, in {\it Model 1} this line turns into first order at a tricritical point.
In \sref{sec:ABC1} we study {\it Model 2} which is shown to exhibit a second order transition line which is different than \eref{eq:bc1}.
Below we define the two grand-canonical generalization by performing the appropriate Legendre transform of $\mathcal{F}$ defined in \eref{eq:abc_energy1}. It is worth noting, however,
that the two models can be studied by considering specific particle-nonconserving dynamical rules in addition to (\ref{eq:abc_rules}) and (\ref{eq:vacancyexchange}) that obey
detailed balance with respect to the ABC Hamiltonian. These dynamical rules are discussed for {\it Model 1} in \cite{Lederhendler2010a,Lederhendler2010b} and for {\it Model 2} in \cite{Barton2010}.

\subsection{Model 1: Grand-canonical ABC model with $r_A-r_B$ and $r_B-r_C$ fixed}
\label{sec:ABC1}

In this section we consider a grand-canonical ABC model where the
 differences between the overall densities of the species, $r_A-r_B$ and $r_B-r_C$, are fixed
while the overall particle density, $r=r_A+r_B+r_C$, is allowed to fluctuate \cite{Lederhendler2010a,Lederhendler2010b}.
We study the model for equal densities, $r_A=r_B=r_C$, which is the only case
where it exhibits a phase transition between the homogenous and ordered phases.
The large deviation function of the density profile is given in this case by
\begin{equation}
\label{eq:abc_gcan_energy}
\mathcal{G}\left[{\boldsymbol \rho}(x),r,\mu\right]=\mathcal{F}\left[{\boldsymbol \rho}(x),r\right]-\beta\mu r,
\end{equation}
where $\mu$ is the chemical potential.
We study the critical point of the model by expanding \eref{eq:abc_gcan_energy} close to the homogenous phase, while allowing for fluctuations in $r$ denoted by $\delta r$. The density profile is thus given by
\begin{eqnarray}
\label{eq:rho_expand1}
{\boldsymbol \rho}(x)=\frac{r+\delta r}{3}{\bf u}+\sum_{n\neq0}\left[b_{n}{\bf v}+b_{-n}^{\star}{\bf v}^\star
\right]e^{2\pi inx},
\end{eqnarray}
where $b_0=0$ since we consider the equal-densities case. In terms of the generic model defined
in \sref{sec:landau_scalar}, it will be shown below that $b_{1}$ and $\delta r$ play the role of $m$ and $\delta q$ , respectively.
Using \eref{eq:rho_expand1} we rewrite the Landau expansion obtained in \cite{Lederhendler2010b} as
\begin{eqnarray}
\label{eq:abc_gcan_energy1}
\mathcal{G}(\{b\})\approx\mathcal{G}_{h}\left(r\right)+\left(\frac{3}{r}-\frac{\sqrt{3}\beta}{2\pi}\right)|b_{1}|^{2}+\left(\frac{3}{r}+\frac{\sqrt{3}\beta}{4\pi}\right)|b_{-2}|^{2}-\frac{3}{r^{2}}|b_{1}|^{2}\delta r\nonumber \\
+\left(\frac{1}{2\left(1-r\right)}+\frac{1}{2r}\right)\left(\delta r\right)^{2}-\frac{3\sqrt{3}}{2r^{2}}(b_{1}^{2}b_{-2}+b_{1}^{\star2}b_{-2}^\star)+\frac{9}{2r^{3}}|b_{1}|^{4}+\mathcal{O}\left(b_{1}^{6}\right),
\end{eqnarray}
where $\mathcal{G}_h\left(r\right)=\mathcal{F}_h\left(r\right)-\beta\mu r$.
The coupling term between the fluctuating field, $\delta r$,
and the primary order parameter, $b_{1}$, in \eref{eq:abc_gcan_energy1} is of the form
$\delta r|b_{1}|^{2}$. Since this is compatible with the generic Landau expansion presented
in \eref{eq:Fdq}, where $f(m,q)=f(-m,q)$, we expect the canonical and grand-canonical ensembles to exhibit the same critical lines.

The form of the coupling between $b_1$ and $\delta r$ can also be deduced from the symmetry properties of the model, without deriving \eref{eq:abc_gcan_energy1} explicitly.
The large deviation function, $\mathcal{F}$, is invariant under cyclic exchange of the species label, $A\to B$, $B\to C$ and $C\to A$, given in terms of the Fourier
modes by
 \begin{equation}
\label{eq:sepcies_exchange}
 b_n \to b_n e^{-2\pi i/3}.
\end{equation}
In the equal-densities case $\mathcal{F}$ can be shown to be invariant also under
a spatial shift of the density profile by a continuous parameter, $\Delta$, which implies that  $\mathcal{F}\left[{\boldsymbol \rho}(x),r\right] = \mathcal{F}\left[{\boldsymbol \rho}(\mathrm{mod}_1(x+\Delta)),r\right] $.
 In terms of the Fourier modes the translation of the profile takes the form of
\begin{equation}
\label{eq:translation}
b_n \to b_n e^{-2\pi i n \Delta}.
\end{equation}
Since the overall particle density, $r+\delta r=\sum_\alpha\int_0^1\rho_\alpha(x)dx$, is invariant under \eref{eq:sepcies_exchange} and \eref{eq:translation},
we expect the lowest order coupling between $b_1$ and $\delta r$ in \eref{eq:abc_gcan_energy1} to be of the form $|b_1|^2\delta r$.

\begin{figure}[t]
\noindent
\begin{centering}\includegraphics[scale=0.7]{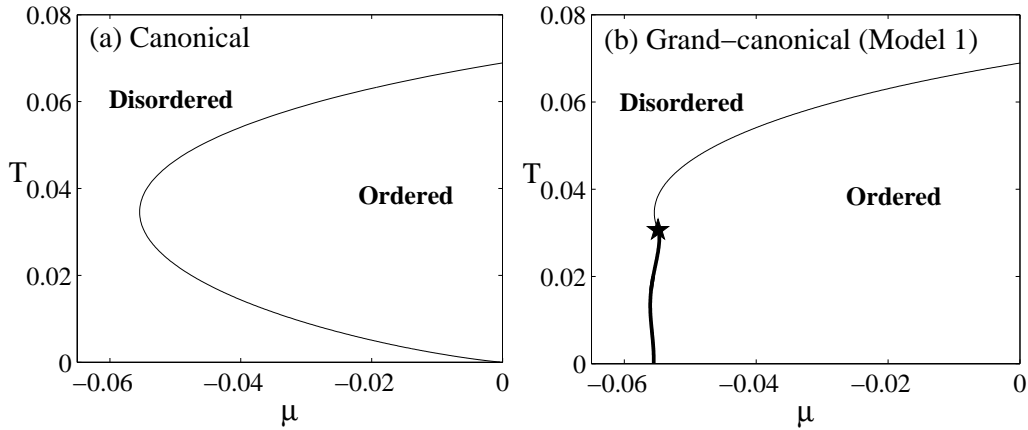}\par\end{centering}
\caption{Phase diagram of the canonical ABC model (a) and {\it Model 1} (b), plotted in the ($\mu,T=1/\beta$) plane.
The two diagrams show the same second order transition line (thin solid line).
In the grand-canonical ensemble the transition line turns into a first order (thick solid line) at a tricritical point ($\star$).
\label{fig:MuT}}
\end{figure}

Our conclusion that the grand-canonical critical line is identical to the canonical one can be demonstrated by calculating the former explicitly.
 To this end, we first minimize $\mathcal{G}$ with respect to $\delta r$ and $b_{-2}$ in order to obtain their expression in terms of the primary order parameter, $b_{1}$,
\begin{equation}
\label{eq:abc_dr_b2}
b_{-2}=\frac{6\pi}{r\left(4\sqrt{3}\pi+\beta r\right)}(b_{1}^\star)^{2},\qquad\delta r=\frac{3\left(1-r\right)}{r}|b_{1}|^{2}.
\end{equation}
Inserting \eref{eq:abc_dr_b2} into \eref{eq:abc_gcan_energy1} yields the following Landau expansion
\begin{equation}
\label{eq:Gabc1}
\mathcal{G}(b_{1})=\mathcal{G}_h\left(r\right)+g_{2}\left(\beta,r\right)|b_{1}|^{2}+g_{4}\left(\beta,r\right)|b_{1}|^{4}+\mathcal{O}\left(|b_{1}|^{6}\right),
\end{equation}
where
\begin{eqnarray}
g_{2}\left(\beta,r\right)=f_{2}\left(\beta,r\right)=\frac{3}{r}-\frac{\sqrt{3}\beta}{2\pi}, \nonumber \\
g_{4}\left(\beta,r\right)=\frac{9}{2r^{3}}\left[\frac{\sqrt{3}\beta r+6\pi}{\sqrt{3}\beta r+12\pi}-\frac{3\left(1-r\right)}{3}\right]\neq f_4\left(\beta,r\right).\label{eq:abc_gcan_energy2}
\end{eqnarray}
We can now compare \eref{eq:Gabc1} to the generic Landau expansion, found in \eref{eq:gmlambda1},
and derive the phase diagram of the grand-canonical ABC model following the lines presented in \sref{sec:landau_scalar}.
On the critical line, $\beta =\beta_c = 2\pi\sqrt{3}/r$, where $g_{2}=0$, we find that
\begin{equation}
g_{4}\left(\beta_{c},r\right)=\frac{3(3r-1)}{2r^{3}},\label{eq:abc_g4}
\end{equation}
which is positive for $r>1/3$. We thus conclude that the canonical and grand-canonical ensembles exhibit the same second order transition line at
\begin{equation}
\beta_{c}=2\pi\sqrt{3}/r, \quad \mathrm{for} \quad r>r_{TCP}=1/3.
\end{equation}
 Below the tricritical point ($r<r_{TCP}$), where $g_4<0$, we find a first order transition line in the grand-canonical ensemble, whereas the canonical transition line remains second order for all $r$. The resulting grand-canonical phase diagram is plotted in the $(\mu,1/\beta)$ plane in \fref{fig:MuT}b.
 The figure shows a second order transition line (thin solid line) which turn into a first order line (thick solid line) at a tricritical point ($\star$).
  The first order transition
line in the figure is obtained from the exact expression of the steady-state density profile of the model, derived in \cite{Lederhendler2010b}.
For comparison we plot the canonical phase diagram in \fref{fig:MuT}a, which is composed only of a second order transition line.
The chemical potential in the canonical ensemble is given by $\mu=d\mathcal{F}_h/dr$.
 As expected, due to the symmetry properties of $\mathcal{F}$, the ensemble inequivalence presented in \fref{fig:MuT} is consistent with the behavior of the generic model studied in section \ref{sec:landau_scalar}.

\subsection{Model 2: Grand-canonical ABC model with $r_A+r_B+r_C$ fixed}
\label{sec:ABC2}
In this section we analyze a different grand-canonical ABC model where $r_A$, $r_B$ and $r_C$ are allowed to fluctuate independently, while maintaining the overall particle density, $r=r_A+r_B+r_C$, fixed.

 The large deviation function of the density profile of {\it Model 2} is given by
\begin{equation}
\mathcal{G}\left[{\boldsymbol \rho}(x),r_\alpha,\lambda_\alpha\right]=\mathcal{F}\left[{\boldsymbol \rho}(x),r\right]-\sum_{\alpha=A,B,C}\lambda_{\alpha}r_{\alpha},
\label{eq:abc_gcan_energy3}
\end{equation}
where $\lambda_\alpha$ are the conjugate fields which control the fluctuations of $r_\alpha$.
Here we study case  $\lambda_{A}=\lambda_{B}=\lambda_{C} \equiv \mu$, for which the model undergoes a phase transition.
 Since $r$ is fixed, changing the value of $\sum_{\alpha}\lambda_{\alpha}$ shifts $\mathcal{G}$ by a constant without affecting the resulting phase diagram.
  The field $\mu$ is thus an arbitrary parameter in this model, which can be set to zero.
We expand $\mathcal{G}$ in terms of small deviations from the homogenous profile, defined as
\begin{equation}
\label{eq:abc_profile1}
{\boldsymbol \rho}(x)=\frac{r}{3}{\bf u}+\sum_{n=-\infty}^\infty\left[b_{n}{\bf v}+b_{-n}^\star{\bf v}^\star
\right]e^{2\pi inx}.
\end{equation}
Unlike in the canonical model, where $b_0$ is fixed, here it is allowed to fluctuate, leading to a significantly different
 phase diagram. To lowest order in $b_n$ we obtain 
\begin{eqnarray}
\label{eq:abc_gcan_enegy7}
\mathcal{G}(\{{b}\}) & = &\mathcal{G}_{h}(r)+( \frac{3}{r}-\frac{\beta}{2})|b_{0}|^{2}+\nonumber \\
& & \sum_{n\neq0}\left[(\frac{3}{r}-\frac{\beta\sqrt{3}}{2\pi n})|b_{n}|^{2}-\frac{\beta\sqrt{3}}{2\pi n}(b_{0}b_{n}^{\star}+b_{0}^{\star}b_{n})\right]+\mathcal{O}(b_{n}^{4}).
\end{eqnarray}
Due to the bilinear coupling between $b_0$ and $b_n$ in $\mathcal{G}$ for arbitrary $n$, {\it all modes have the same symmetry} and any one of them can be viewed as the primary order parameter. As shown below, there is a specific combination of all the modes which becomes linearly unstable at the critical point.
The bilinear coupling term suggests that this model is compatible with
the generic Landau expansion in \eref{eq:Fdq2}, where $f(m,q)=f(-m,-q)$, with $b_1$ and $b_0$ playing
the role of $m$ and $q$, respectively. In {\it Model 2}, however, we find an additional infinite set of parameters, $b_n$ for $n\neq 0,1$,
which are coupled linearly to $b_0$ and thus play a role in setting the critical point of the model \cite{canonical_note}.
Even though \eref{eq:abc_gcan_enegy7} has multiple order parameters coupled linearly to $b_0$
and each of them is complex, the resulting behaviour should be similar to that of the generic model analyzed in \sref{sec:landau_tensor}.
This implies that the canonical and grand-canonical ensembles should exhibit different critical points, as demonstrated explicitly below.

The form of the coupling term between $b_0$ and $b_1$ in \eref{eq:abc_gcan_enegy7} can be deduced without performing the Landau expansion explicitly, simply
 by analyzing the symmetry properties of the model.
 Here, since we cannot restrict the system to remain in the equal-densities point,
 the translational symmetry is broken and $\mathcal{F}$ is invariant only
under cyclic exchanges of the species' labels given by
\begin{equation}
b_n \to b_n e^{-2\pi i/3}.
\end{equation}
This implies that the Landau expansion of $\mathcal{F}$ may contain bilinear coupling terms of the form  $b_{0}b_{n}^\star$ and  $b_{0}^\star b_{n}$.
In principle, from these symmetry considerations, we also expect to find terms of the form $b_{n}b_{l}^{\star}$ with $n,l\neq0$. However, explicit calculations show that the coefficients of these terms vanish.

We now demonstrate that the two ensembles exhibit different critical lines by computing the grand-canonical one explicitly.
Since each of the Fourier modes can be regarded as the primary order parameter we choose to express them in terms of $b_0$ and then
obtain an expansion of $\mathcal{G}$ in powers of $b_0$.
This can be done by first minimizing $\mathcal{G}$ with respect to $b_n$ for $n\neq0$ which yields
\begin{equation}
\label{eq:abc_bm}
\frac{\partial\mathcal{G}}{\partial b_{n}^{\star}}=(\frac{3}{r}-\frac{\beta\sqrt{3}}{2\pi n})b_{n}-\frac{\beta\sqrt{3}}{2\pi n}b_{0}=0,
\end{equation}
and thus
\begin{equation}
\label{eq:gcan_bm}
b_{n}=\frac{1}{\frac{2\pi n\sqrt{3}}{\beta r}-1}b_{0}.
\end{equation}
 We insert \eref{eq:gcan_bm} this into \eref{eq:abc_gcan_enegy7} and obtain a Landau expansion of $\mathcal{G}$ in terms of $b_0$,
\begin{equation}
\label{eq:abc_b0_series}
\mathcal{G}(b_0) = \mathcal{G}_{h}(r)+\left[( \frac{3}{r}-\frac{\beta}{2}) -\frac{\beta\sqrt{3}}{2\pi}\sum_{n\neq0}\frac{1}{\frac{2\pi\sqrt{3}}{\beta r}n^{2}-n}\right]|b_{0}|^2+O(|b_{0}|^4).
\end{equation}
By simplifying the coefficient of $|b_0|^2$ above, as detailed in \ref{app:bc}, it can be shown to vanish at
\begin{equation}
\label{eq:bc_interval}
\beta r =2\pi\sqrt{3}(l+\frac{1}{3}) \qquad l \in \mathbb{Z}.
\end{equation}
The highest temperature, $T=1/\beta$, for which the instability sets in is thus obtained for $l=0$, yielding
\begin{equation}
\label{eq:model2_bc}
\beta_c=\frac{2\pi\sqrt{3}}{3r}.
\end{equation}
At this point $b_0$ becomes unstable along with the rest of the modes which depend linearly on $b_0$ through \eref{eq:gcan_bm}.
This critical point has been derived in \cite{Barton2010} by computing the exact minimizing density profile of the large deviation function of {\it Model 2} (\ref{eq:abc_gcan_energy3}).

\begin{figure}[t]
\noindent
\begin{centering}\includegraphics[scale=0.7]{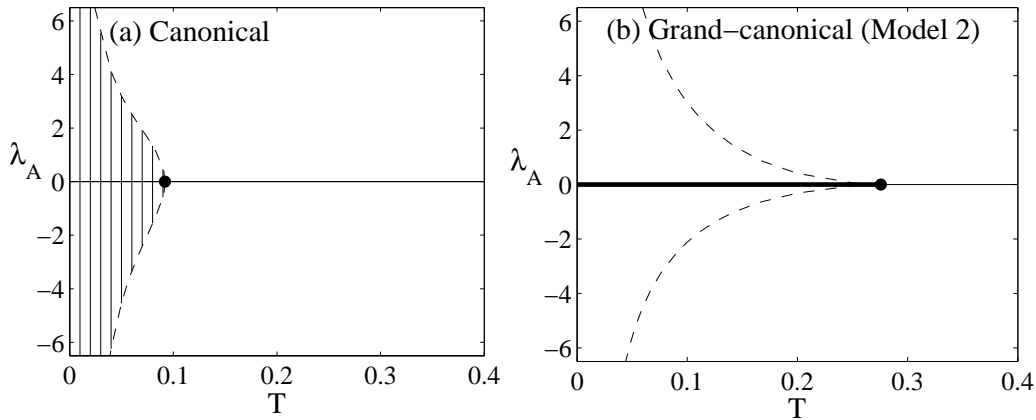}\par\end{centering}
\caption{
Phase diagram of the canonical ABC model (a) and {\it Model 2} (b), plotted in the ($\lambda_A,T=1/\beta$) plane for the case where
$\lambda_A=\lambda_B=-\lambda_C/2$ and $r=1$.
Here, the two ensembles exhibit different critical points ($\bullet$). Below the critical points we find a first order transition,
denoted in (a) by a discontinuity in $\lambda_A$ (vertical lines) and in (b) by a thick solid line.
The dashed lines in (b) mark the stability limits of the meta-stable phases, computed using the exact solution of the density profile.
 They define a coexistence region where a two different ordered profiles are locally stable in {\it Model 2}. See details in text.
\label{fig:MuT_interval}}
\end{figure}

As expected, since the Landau expansion involves a bilinear coupling term between the fluctuating field, $b_0$, and the order parameter of the canonical model, $b_1$,
the canonical and grand-canonical models exhibit different
critical points at $\beta=2\pi\sqrt{3}/r$ and $\beta=2\pi\sqrt{3}/3r$, respectively. The phase diagrams, derived within the two ensembles, are shown in \fref{fig:MuT_interval}.
 The figure corresponds to the two-equal densities case, $r_A=r_B\neq r_C$, obtained by setting $\lambda_A=\lambda_B=-\lambda_C/2$. In the canonical ensemble we find a second order phase transition
point ($\bullet$) on the $\lambda_A=0$ line. Below this point the model undergoes a first order
phase transition between  a $C$-rich ($r_A=r_B<r_C$) and a $C$-poor ($r_A=r_B>r_C$) ordered phases, depicted
by the discontinuity in $\lambda_A=d\mathcal{F}/d r_A$ (vertical lines in \fref{fig:MuT_interval}a).
  In the grand-canonical
 phase diagram, shown in \fref{fig:MuT_interval}b, the second order phase transition ($\bullet$) is found at a higher temperature. Below the critical temperature the model undergoes a first order phase transition at $\lambda_A=0$ between a $C$-rich and a $C$-poor ordered phases (thick solid line). The $C$-rich ($C$-poor)
 phase is meta-stable for $\lambda_A>0$ ($\lambda_A<0$) up to the dashed line where it becomes unstable. The stability limits (dashed lines)
 are obtained using the exact steady-state density profile of the ABC model, derived in \cite{Ayyer2009,Barton2010}.

\section{Conclusions}
\label{sec:conc}
In this paper we relate
the order of the phase transition where two ensembles of a long-range interacting system may become inequivalent
to the symmetry of the underlying order parameter of the transition.
This is done within the framework of Landau theory, which constitutes a general approach for studying
systems with long-range interactions.

For a transition governed by an order parameter $m$ and a field $q$ which is allowed to fluctuate only
in the higher ensemble, we consider two
types of symmetries of the Landau free energy density, $f(m,q)$. In the first case, where $f(m,q)=f(-m,q)$,
 the lowest order coupling between $m$ and $q$ is of the form $m^2 q$, yielding fluctuations in $q$
 that scale as $\delta q = q- q^\star\sim m^2$. These fluctuations cannot affect
 the critical line of the model which depend on the coefficient of $m^2$ in the Landau expansion of the model.
We thus find that the two ensembles, where $q$ is either fixed or allowed to fluctuate,
 differ only in their first order transition lines and tricritical points which depend on higher order terms
 in the Landau expansion. In the second case, we consider a symmetry of the form $f(m,q)=f(-m,-q)$ and analyze the stability of the phase where $m^\star=q^\star=0$.
  Here, the Landau expansion may involve a bilinear coupling term of the form $m q$, yielding fluctuations in $q$ that scale as $\delta q\sim m$.
These fluctuations affect the coefficient of the $m^2$ term in the Landau expansion,
causing the two ensembles to exhibit different critical lines.

An example for the first symmetry, $f(m,q)=f(-m,q)$, can be found when considering $m$ to be the average magnetization of the system and $q$ to be the energy or the overall density, which are both invariant under $m \to -m$. Since this is the most natural choice for $q$, our derivation explains why most of the
examples of ensemble inequivalence that were studied in the past are of this type (see e.g. \cite{Barre2002,Misawa2006,Ellis2004,Mukamel2005,Lederhendler2010a,Lederhendler2010b,Grosskinsky2008}).
The second symmetry, $f(m,q)=f(-m,-q)$, is shown above
to exist in an anisotropic mean-field XY model as well as in the generalized ABC model studied in \cite{Barton2010}.
We provide detailed examples of the two different symmetries, by deriving the Landau expansion of two
distinct grand-canonical generalizations of the ABC model.
Our general analysis is found to be applicable for the ABC model, even though the latter has
a complex order parameter and a three-fold symmetry.

We thus conclude that the symmetry of the model sets the order of transition where ensemble
inequivalence may be observed. If the symmetry permits a bilinear coupling term in the Landau expansion
 between the order parameter which governs the transition in the lower ensemble, $m$, and the parameter which is allowed to fluctuate in the higher ensemble, $q$, the two ensembles would generically exhibit different second order transition points.

\ack We thank A. Bar, O. Hirschberg, J.L. Lebowitz, T. Sadhu and E.R. Speer for helpful discussions. The support of the Israel Science
Foundation (ISF) is gratefully acknowledged.

\appendix
\section{Derivation of the critical point of Model 2}
\label{app:bc}
We wish to find the critical point of {\it Model 2}, described in section \ref{sec:ABC2}.
To this end we obtained the Landau expansion of the model in powers of $b_0$, given by
\begin{equation}
\label{eq:abc_b0_series2}
\mathcal{G}(b_0) = \mathcal{G}_{h}(r)+ \left[( \frac{3}{r}-\frac{\beta}{2}) -\frac{\beta\sqrt{3}}{2\pi}\sum_{n\neq0}\frac{1}{\frac{2\pi\sqrt{3}}{\beta r}n^{2}-n}\right]|b_{0}|^2+O(|b_{0}|^4).
\end{equation}
The critical point is found where the coefficient of $|b_0|^2$ vanishes. This coefficient can be simplified by expressing it in terms of the DiGamma function,
 defined as
\begin{equation}
\psi(1+y)=-\gamma+\sum_{k=1}^{\infty}\left(\frac{1}{k}-\frac{1}{k+y}\right)=
-\gamma+\sum_{k=1}^{\infty}\left(\frac{1}{\frac{k^2}{y}+k}\right),\label{eq:digamma}
\end{equation}
where $\gamma$ is the Euler-Mascheroni constant.
Using the recursive properties of the DiGamma function given in \cite{Abramowitz64} (section 6.3),
\begin{equation}
\psi(1-y)=\psi(y)+\pi\cot(\pi y),
\end{equation}
and
\begin{equation}
\psi(1+y)=\psi(y)+\frac{1}{y},
\end{equation}
the sum over $n$ in \eref{eq:abc_b0_series2} can be written as
\begin{equation}
\label{eq:digamma_eq1}
\sum_{n\neq0}\frac{1}{n^{2}/y-n}=\psi(1+y)-\psi(1-y)=\frac{1}{y}-\pi\cot(\pi y),
\end{equation}
where $y=\beta r/2\pi\sqrt{3}$. \Eref{eq:abc_b0_series2} can thus be written as
\begin{eqnarray}
\mathcal{G}(\{{b}\}) &=& \mathcal{G}_{h}(r) +\\
&&( \frac{3}{r}-\frac{\beta}{2})\left[1-\left( \frac{1}{y}-\frac{\pi}{\sqrt{3}} \right)^{-1}\left(\frac{1}{y}-\pi\cot{\pi y}\right)\right]|b_{0}|^2+O(|b_{0}|^4) \nonumber.
\end{eqnarray}
The critical point is found where the coefficient of $|b_0|^2$ vanishes, yielding the equation
\begin{equation}
\frac{1}{y}-\pi\cot{\pi y} =\frac{1}{y}-\frac{\pi}{\sqrt{3}},
\end{equation}
which can be written as
\begin{equation}
\label{eq:cotang}
\cot\left(\frac{\beta r}{2\sqrt{3}}\right)=\frac{1}{\sqrt{3}}.
\end{equation}
\Eref{eq:cotang} has several solutions given by
\begin{equation}
\beta r =2\pi\sqrt{3}(l+\frac{1}{3}) \qquad l \in \mathbb{Z}.
\end{equation}
 The solution for $l=0$ has the highest temperature, $T=1/\beta$, and thus corresponds to the critical point of the model,
\begin{equation}
\beta_{c}=\frac{2\pi\sqrt{3}}{3 r}.
\end{equation}
This point was derived in \cite{Barton2010} by the studying the exact expression for the steady-state density profile of the model.
Note that the critical point is different than the canonical one, given by $\beta_{c}=\frac{2\pi\sqrt{3}}{ r}$. They differ
by a factor of $3$, whose source is discussed in detail in \cite{Barton2010}.

\bibliographystyle{iopart-num}
\bibliography{ABCModel}

\begin{thebibliography}{10}
\bibitem{canonical_note}
In the canonical ensemble, where $b_0=0$, these additional parameters are
  determined by their higher order coupling terms to $b_1$, which yield the
  scaling given in \eref{eq:abc_b_scaling}. They therefore do not affect the
  critical point of the canonical model, set by the coefficient of $|b_1|^2$ in
  the Landau expansion.
\end{thebibliography}

\end{document}